\documentclass[aps,prb,twocolumn,showpacs,superscriptaddress]{revtex4-1}

\usepackage{graphicx}
\usepackage{dcolumn}
\usepackage{bm}
\usepackage{amsmath} 
\usepackage[dvips]{color}

\begin{document}

\title{Binding carriers to a non-magnetic impurity in a
  two-dimensional square Ising antiferromagnet}  

\author{Hadi Ebrahimnejad} 
\affiliation{Department of Physics and Astronomy, University of
  British Columbia, Vancouver, BC, Canada, V6T 1Z1}

\author{Mona Berciu} 
\affiliation{Department of Physics and Astronomy, University of
  British Columbia, Vancouver, BC, Canada, V6T 1Z1} 
\affiliation{Quantum Matter Institute, University of British Columbia,
Vancouver, BC, Canada, V6T 1Z4}

\date{\today}

\begin{abstract}
A hole in a two-dimensional Ising antiferromagnet was believed to be
infinitely heavy due to the string of wrongly oriented spins it
creates as it moves, which should trap it near its original
location. Trugman showed that, in fact, the hole acquires a finite
effective mass due to contributions from so-called {\it Trugman loop}
processes, where the hole goes nearly twice around closed loops, first
creating and then removing wrongly-oriented spins, and ending up at a
different lattice site.  This generates effective second- and third-nearest-neighbor hopping terms which keep the quasiparticle on the
sublattice it was created on. Here, we investigate the trapping of the
quasiparticle near a single attractive non-magnetic impurity placed 
at one lattice site. We consider the two cases with the quasiparticle
and impurity being on the same versus on different sublattices. The
main result is that even 
though the quasiparticle can not see the bare disorder in the latter
case, the coupling to magnons generates an effective renormalized
disorder on its own sublattice which is strong enough to lead to bound
states, which however have a very different spectrum than when the
quasiparticle and impurity are on the same sublattice.
\end{abstract}

\pacs{75.50.Ee, 74.72.Gh, 71.23.An}
\maketitle

\section{introduction}
Understanding the motion of charge carriers in a two-dimensional (2D)
Heisenberg antiferromagnet (AFM) is a central challenge for
deciphering the mechanism behind high-temperature superconductivity in
cuprates.\cite{Kane, Frank, Sachdev, Chubukov, Sigia} At half-filling,
the strong hybridization $t_{pd}$ between copper $d_{x^2-y^2}$ and
oxygen $p_{x,y}$ orbitals drives the CuO$_2$ planes into an insulating
state in which the holes on neighboring copper atoms align their
spins antiferromagnetically in order to gain the superexchange energy
$\mathrm{J}\sim t_{pd}^4/\Delta^3$, where $\Delta=E_p-E_d$ is the
charge transfer energy from $d$ to $p$ orbitals. Superconductivity
emerges upon doping the 2D AFM planes with charge carriers.\cite{Lee, Phillips}  

A major setback in the search for an analytic description of the
behavior of these charge carriers is the lack of a simple wave
function for the ground state of the undoped AFM planes. The
semi-classical N{\'e}el state breaks  spin rotation
symmetry and is therefore smeared out by quantum spin
fluctuations  to a significant degree that is hard to capture with
simple wave functions. This leaves numerical calculations as the only way to
make quantitative predictions.\cite{Dagotto} While implementing such
numerical calculations is already a complicated task even for a clean
system, a further complication comes from the presence of disorder and
imperfections in the real materials, introduced during the sample
growth and preparation. Given the low dimensionality, even weak
disorder may have dramatic effects on the motion of charge carriers in the
CuO$_2$ planes. It is well known that non-magnetic impurities are strong pair 
breakers in $d$-wave superconductors.\cite{Riera0}
Indeed, substitution of only a few percent of the
copper atoms with non-magnetic impurities has been observed to suppress the
superconductivity by localizing the low-energy electronic
states.\cite{basov} Even in the cleanest samples, the dopant ions are
in close proximity to the CuO$_2$ planes, and the disorder potential they create
can disturb the motion of the charge carriers.  

Impurities have been shown to be responsible for a range of phenomena
in low-dimensional correlated electron systems, and they can be also
utilized for probing correlations which are otherwise difficult to
observe in the ground state.\cite{Alloul} For the undoped parent
compound, mean-field analysis of the disordered Hubbard model predicts
the emergence of an inhomogeneous metallic phase in which the Mott gap
is locally closed wherever the disorder is strong enough to do
so.\cite{Heidarian} However, it is not always the case that impurities
destroy the order in the underlying system. For instance, impurities
induce local magnetic order in one-dimensional (1D) quantum magnets,\cite{Suchaneck} and
long-range antiferromagnetism is predicted upon doping some quantum
spin liquids with nonmagnetic impurities.\cite{wessel}
In any event, a
complete understanding of the interplay between disorder and AFM
correlations and especially of their role in controlling the carrier
dynamics away from half-filling is still lacking.

In this paper, we consider a much simpler variant of this problem
where, at zero temperature, a hole is created in a 2D Ising AFM on a
square lattice, and is also subject to the on-site attractive potential of an
impurity that can be visited by the hole.  Thus, our
model is very different from  previous models of an impurity in a 2D Heisenberg
AFM, which assumed that the hole cannot visit the impurity
site, and is coupled to it at most through
exchange.\cite{Poilblanc1,Poilblanc2} As we discuss in the following, our results 
have some similarities but also considerable differences from those
obtained numerically in these other models.

We investigate the local density
of states (LDOS) near the impurity to study the appearance of bound
states, focusing specifically on the relevance of the magnetic
sublattice on which the impurity is located. The advantage of our
approach is that the wave function of the undoped 2D AFM is the simple
N{\'e}el state, and this allows us to study the problem
(quasi)analytically. Of course, spin fluctuations are completely absent, but,
as we argue in our discussion, our results allow us to speculate about
(at least some of) their likely effects.

We note that a single hole in an Ising AFM was initially believed to be
localized even in the absence of impurities, because when the hole hops
it reshuffles the spins along its path, thus creating a string of wrongly
oriented spins. In dimensions larger than one, the
energy cost of this string increases roughly linearly with its length,
resulting in an effective potential well that binds the hole in
the vicinity of its original position. Finite mobility was
believed to arise only due to spin fluctuations which can
remove pairs of such defects,\cite{Brink, Poilblanc3} but they are
absent from the Ising Hamiltonian.  

However, as pointed out by Trugman,\cite{Trugman} the hole is
actually delocalized even in the Ising AFM, and it achieves this
by going twice around closed loops. The string of misaligned spins
that are created in the first round is removed when spins are
reshuffled again during the second round. When the last one is
removed, the hole ends up at a different site from where it started,
and by repeating this process it can move anywhere on its original
sublattice (spin conservation ensures that the hole
propagates on one sublattice). This raises the question of
how the hole's motion will be affected by an attractive impurity, especially by
one located on the other sublattice than the one on which the hole resides. 
While one  expects  the hole to become bound to the impurity if they
are on the same sublattice, if they are on different sublattices, one
may expect the hole not to be sensitive to the presence of the
impurity and therefore remain unbound.  We investigate this problem
using a variational method introduced in Ref. \onlinecite{Mona} to
study the clean case, which we generalize here to systems
that are not invariant under translations. Our results confirm the
expected existence
of a bound state when the hole and impurity are on the same
sublattice. When they are on different sublattices, we find
that the naive picture described above is wrong: the hole develops multiple
bound states with a characteristic spectrum and symmetries. The
implications of these results as seen in the
wider context of the effect of disorder on
dressed quasi-particles are also discussed.

This paper is organized as follows: we introduce the model in Sec.
\ref{sec:model}. The generalization of the variational method to
inhomogeneous systems is discussed in Sec. \ref{sec:clean}, followed
in Sec. \ref{sec:disordered} by results for a single impurity located
(i) on the same, and (ii) on the other sublattice than the hole. We conclude the
paper by giving a summary and discussing possible further developments
of this work in Sec. \ref{sec:symmary}.

\section{model}
\label{sec:model}
We consider the motion of a single hole doped into a spin-$1/2$ Ising
antiferromagnet on a 2D square lattice. The Hamiltonian of the undoped
system is  
\begin{equation}
{\cal{H}}_\mathrm{AFM}=\mathrm{J}\sum_{<i,j>}[S_i^zS_j^z+\frac{1}{4}]=
\mathrm{\bar{J}}\sum_{<i,j>}[\sigma_i^z\sigma_j^z+1], 
\end{equation}
where $\sigma^z$ is the Pauli matrix and $\mathrm{\bar{J}}=\mathrm{J}/4>0$.
The vacuum $|0\rangle$ is the N{\'e}el-ordered state, with all spins
on one sublattice pointing up and those on the other sublattice
pointing down. Excitations are gapped spin-flips, or localized
magnons, and we refer to them also as spin defects. The creation operator
for a spin defect is written in terms of the spin raising and lowering
operators, $\sigma^{\pm}=\sigma^x\pm i\sigma^y$: 
\begin{equation} 
d^{\dagger}_i=
  \begin{cases}
   \sigma_i^- & \text{if\,\,\,\,\, } i\in \uparrow \mathrm{sublattice}, \\ \sigma_i^+ &
   \text{if\,\,\,\,\, } i\in \downarrow \mathrm{sublattice}.
  \end{cases}
\end{equation} 

Consider now the doped case. Creating a hole in this system corresponds to removing a spin from the
same lattice site, therefore the hole creation operators are: 
\begin{equation}
h^{\dagger}_i=
  \begin{cases}
   c_{i\uparrow} & \text{if\,\,\,\,\, } i\in \uparrow \mathrm{sublattice}, \\ c_{i\downarrow} &
   \text{if\,\,\,\,\, } i\in \downarrow \mathrm{sublattice}.
  \end{cases}
\end{equation}  
Once the hole is created ($h_i^{\dagger}$), it moves via nearest-neighbor hopping. This, however,
either creates a spin defect on the hole's departure site ($d^{\dagger}_i$)
or annihilates one from its arrival site ($d_j$), if there was a spin
defect already there. The Hamiltonian can therefore be written as:\cite{Mona} 
\begin{equation}
\label{H}
{\cal{H}}={\cal{P}}\{-t\sum_{<ij>}[h_j^{\dagger}h_i(d^{\dagger}_i+d_j)+\mathrm{H.c.}]\}
{\cal{P}}+{\cal{H}}_\mathrm{AFM}-Uh^{\dagger}_0h_0,
\end{equation}
where ${\cal{P}}$ is the projection operator enforcing no double
occupancy: at any site there is a hole or there is a
spin which is either properly oriented or is flipped,
$h_i^{\dagger}h_i+d_i^{\dagger}d_i+d_id_i^{\dagger}=1$.  Thus, the first
term describes the hopping of the hole which is accompanied by either
spin defect creation or annihilation. 

In addition, there is an attractive  potential  of strength $U$
centered at the origin  $\bf r=0$,  
which changes the on-site energy of the visiting hole (variations of
the local hoppings and exchanges can be trivially included in the
model and our solution, but should not lead to any qualitative
changes if they are small or moderate in size). Physically, 
such a potential can be due to an attractive non-magnetic impurity located above
the origin, in a different layer, and which  modulates the
on-site energy at the origin. Another possibility comes from  replacing
the atom at the origin by an impurity atom with the same valence, but
whose orbitals lie at lower energies than
those of the background atoms.  This is very different from the
impurity models studied in previous
work where the impurity is an inert site that can not be visited by carriers,
\cite{Poilblanc1} and there is at most exchange between the spin of
the  impurity and that of carriers located on neighboring sites.\cite{Poilblanc2}

\section{Propagation of the hole in the clean system}
\label{sec:clean}
In this section, we construct the equations of motion for the
zero-temperature Green's function (GF) of a single hole moving through
the lattice in the absence of impurity, $U=0$. This was done
using a momentum-space formulation in Ref. \onlinecite {Mona}. Here, we present a
real-space derivation, whose use becomes inevitable once we introduce
the impurity which breaks the translational invariance. The single
hole GF is defined as  
\begin{equation}
\label{G0R0}
G_{\bf 0,R}(\omega)=\langle \mathrm{0}|h_{\bf 0}\hat{G}(\omega)h^{\dagger}_{\bf R}|\mathrm{0}\rangle,
\end{equation}     
where $\hat{G}(\omega)=\lim_{\eta\rightarrow 0^+} 1/(\omega-{\cal
  H}+i\eta)$ is the resolvent associated with the Hamiltonian when
$U=0$.  

By dividing the Hamiltonian as ${\cal H}={\cal H}_\mathrm{AFM}+{\cal
  H}_t$ where ${\cal H}_t$ is the first term in Eq. (\ref{H})
responsible for hopping, equations of motion for $G_{\bf 0,R}(\omega)$
can be generated by repeated use of the Dyson
identity $$\hat{G}(\omega)=\hat{G}_\mathrm{AFM}(\omega)+\hat{G}(\omega){\cal
  H}_t\hat{G}_\mathrm{AFM}(\omega),$$ in which
$\hat{G}_\mathrm{AFM}(\omega)=\lim_{\eta\rightarrow 0^+}
1/(\omega-{\cal H}_\mathrm{AFM}+i\eta)$. Using this, Eq. (\ref{G0R0})
becomes 
\begin{equation}
\label{GR}
G_{\bf 0,R}(\omega)=g_0(\omega)[\delta_{\bf 0,R}-t\sum_{\bf u}F_1(\bf R, u;\omega)],
\end{equation}   
where $g_0(\omega)=1/(\omega-4\mathrm{\bar{J}}+i\eta)$ and
$4\mathrm{\bar{J}}$ is the cost of breaking four AFM bonds when
introducing the hole in the lattice. Here the lattice constant is set to
unity, $a=1$, ${\bf u=\pm x, \pm y}$ is any of the four
nearest-neighbor vectors and $F_1({\bf R}, {\bf u}; \omega)=\langle
0|h_{\bf 0}\hat{G}(\omega)d^{\dagger}_{\bf R}h^{\dagger}_{\bf R +
  u}|0\rangle$ has the hole at a nearby site ${\bf R+u}$ and a spin
defect at ${\bf R}$. To simplify notation, from now on we do not write
explicitly the dependence on $\omega$ of all these GFs.  

The equation of motion for $F_1$ can be similarly generated. Upon
application of the Dyson identity, the hole can hop back to ${\bf R}$
and remove the spin defect, or it can hop further away and create a
second spin defect, with an associated GF $F_2$, and so on. As
discussed, states with many spins defects are  less 
likely to occur due to the energy cost of creating the spin defects. In
order to avoid the rise in the number of spin defects, the hole can
trace back its path to remove the spin defects, however this
effectively confines the hole to the vicinity of its creation
site. The hole is freed to move on the lattice
by the so-called {\it Trugman} loop processes in which it goes twice
around a closed path. In this case, spin defects that are created at
the first pass are annihilated when the hole arrives there the second
time. Furthermore, when the very last spin defect is annihilated, the
hole ends up two hops away from its starting point, which is
equivalent to either second- or third-nearest-neighbor hopping
on the main lattice (i.e., first- or second-nearest-neighbor hopping on the
hole's sublattice).

\begin{figure}[t]
\includegraphics[width=0.9\columnwidth]{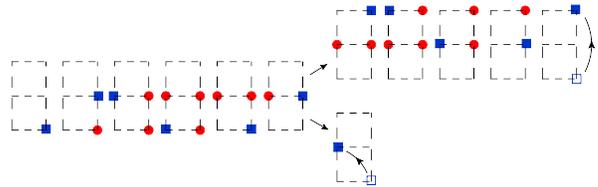}
\caption{Effective first- and second-nearest-neighbor hoppings
  of the hole (the blue square) achieved with loops involving only up to
  three spin defects. The latter is realized when the hole starts a
  second loop before removing the last spin defect it created during
  the first loop. The spin defects are shown
  by red circles. The properly oriented spins are not shown.
}  
\label{hoping1}
\end{figure}      

Longer loops involve more costly intermediate states with more spin
defects, therefore we can proceed within a variational approach in
which a limit is set for the maximum number of spin defects that can
be generated as the hole propagates. We choose to work with up to
three spin-defects, which is the minimum number necessary for the hole
to complete the shortest possible loop. Moreover, we only keep spin
defect configurations consistent with these short closed loops
(i.e., we exclude, for example, configurations where all three spin
defects are collinear). Figure \ref{hoping1} shows how
both types of effective hoppings can be generated with the
three spin defects types of configurations that we keep in our
variational calculation. One can include more configurations  with numerical
simulations, but this was shown to results in only quantitative
differences as long as $t/\mathrm{J}$ is not too
large.\cite{Riera,Mona}

Coming back to the equation of motion for $F_1$, it relates it to
$G_{\bf 0,R}(\omega)$ and also to three GSs $F_2$ with two spin
defects. One of these, with the two spin defects collinear with the hole,
 cannot lead to a closed loop without generating more than
three spin defects, therefore we exclude it from the variational
space, as discussed. Hence, we are left with only three terms 
\begin{equation}
\label{F1}
F_1({\bf R, u})=-tg_1[G_{\bf 0, R}+\sum_{\bf v\perp u}F_2({\bf R, u, v})],
\end{equation}      
where $F_2({\bf R},{\bf u},{\bf v})=\langle 0|h_{\bf
  0}\hat{G}(\omega)d^{\dagger}_{\bf R}d^{\dagger}_{\bf
  R+u}h^{\dagger}_{\bf R + u+v}|0\rangle$, ${\bf v}=\pm {\bf x}$ if
${\bf u} \in \{ {\bf y, -y}\}$ and vice versa,  
$g_1=1/(\omega-10\mathrm{\bar{J}}+i\eta)$ and $10\mathrm{\bar{J}}$ is
the cost of having a spin defect near the hole. 

Within our variational space, the equation of motion for $F_2$ is
\begin{equation}
\label{F2-3}
F_2({\bf R, u, v})=-tg_2[F_1({\bf R, u})+F_3({\bf R, u, v,-u})],
\end{equation}
where 
$$F_3({\bf R},{\bf x},{\bf y},{\bf z})=\langle 0|h_{\bf
  0}\hat{G}(\omega)d^{\dagger}_{\bf R}d^{\dagger}_{\bf
  R+x}d^{\dagger}_{\bf R + x+y}h^{\dagger}_{\bf R+x+y+z}|0\rangle$$
and $g_2=1/(\omega-14\mathrm{\bar{J}}+i\eta)$, where
$14\mathrm{\bar{J}}$ is the cost of the allowed two spin-defects
configurations. The other two three spin-defect configurations
that can be reached starting from $d^{\dagger}_{\bf
  R}d^{\dagger}_{\bf R+u}h^{\dagger}_{\bf R + u+v}|0\rangle$ do not
belong to our variational space and are discarded. Finally, in this
variational space  $F_3$ relates to
$F_2$ only 
\begin{multline}
\label{F3}
F_3({\bf R},{\bf u},{\bf v},{\bf -u})=\\
-tg_3[F_2({\bf R, u, v})+F_2({\bf R+u+v, -v, -u})],
\end{multline}
with $g_3=1/(\omega-16\mathrm{\bar{J}}+i\eta)$, $16\mathrm{\bar{J}}$
being the energy of the allowed three-spin-defect configurations.

These equations can be used to eliminate all $F_3, F_2, F_1$ unknowns and
be left with equations involving  only $G_{\bf 0,R}(\omega)$. The
details are presented in the Appendix. The final results is:
\begin{multline}
\label{GRG}
G_{\bf 0, R}(\omega)=\bar{g}_0(\omega)[\delta_{\bf R,0}\\
-t_1(\omega)\sum_{{\bf \boldsymbol\delta}}G_{\bf 0,
  R+{\boldsymbol\delta}}(\omega)-t_2(\omega)\sum_{{\boldsymbol
    \xi}}G_{\bf 0, R+\boldsymbol\xi}(\omega)], 
\end{multline}
in which
$\bar{g}_0(\omega)=1/(\omega-4\mathrm{\bar{J}}+4t\zeta_1(\omega)+i\eta)$,
$t_1(\omega)=2t\zeta_3(\omega)$, $t_2(\omega)=t\zeta_2(\omega)$ and
$\boldsymbol\delta={\bf\pm u \pm v}$ and $\boldsymbol\xi={\pm2\bf u}$ are
all the second- and third-nearest-neighbor vectors of the full
lattice, respectively. The
explicit expressions of the $\zeta$ functions are given in the
Appendix. 

Equation (\ref{GRG}) shows that the motion of the hole is similar to that
of a quasiparticle with effective second- and third-nearest-neighbor
hoppings $t_1(\omega)$ and $t_2(\omega)$, respectively, and an
effective on-site energy
$\varepsilon(\omega)=4\mathrm{\bar{J}}-4t\zeta_1(\omega)$.  This quasiparticle is
comprised of the hole accompanied by a cloud of spin defects which are
constantly created and annihilated, helping to release the
quasiparticle to move freely on the lattice. Note that all sites ${\bf R }$, ${\bf
  R+\boldsymbol \delta}$, and ${\bf R+\boldsymbol\xi}$ belong to the
same sublattice. Therefore, the quasiparticle propagates on the
sublattice on which the hole is originally introduced, and for which
$\boldsymbol\delta$ and $\boldsymbol\xi$ are the first- and second-nearest-neighbor vectors, respectively.  The constraint that keeps
the quasiparticle moving on one sublattice is very general, being due
to the spin-conserving nature of the Hamiltonian which prevents the
hole from ending up on the other sublattice in the absence of spin
defects: if the hole starts on one sublattice and ends up on the other
one, the $z$ component of the total spin angular momentum of the
system changes from $S^z_i=\pm 1/2$ to $S^z_f=\mp 1/2$, therefore
there needs to be an odd number of spin defects around to compensate
for the change of spin $S_f^z-S_i^z=\mp 1$.

Before presenting the real-space solution of Eq. (\ref{GRG}), note
that we are now in  a position to construct the momentum-space Green's
function: 
\begin{equation}
G({\bf k};\omega)=\langle \mathrm{0}|h_{\bf
  k}\hat{G}(\omega)h^{\dagger}_{\bf
  k}|\mathrm{0}\rangle=\frac{1}{\omega+i\eta-\epsilon(\omega;{\bf
    k})}, 
\end{equation}
where $h_{\bf k}=\sum_{\bf r}\exp(-i{\bf k\cdot{\bf r}})h_{\bf
  r}/\sqrt{\bar{N}}$ and the sum is over the sites in the hole's
sublattice and $\bar{N}\rightarrow \infty$ is their
number. $\epsilon(\omega;{\bf k})$ is the self-energy coming from
coupling to the spin degrees of freedom, which is responsible for the
dynamical generation of the hole's energy dispersion: 
\begin{multline}
\label{Gk}
\epsilon(\omega;{\bf k})=\varepsilon(\omega)-2t_1(\omega)[\cos(k_x+k_y)+\cos(k_x-k_y)]\\
-2t_2(\omega)[\cos(2k_x)+\cos(2k_y)].
\end{multline} 
As required, this is identical to the solution derived using a
momentum space formalism in Ref. \onlinecite{Mona}.\cite{typo}
The spectral function $A({\bf k};\omega)=-\mathrm{Im}G({\bf
  k};\omega)/\pi$  is then used to identify the quasiparticle
excitations and their various properties such as energy
dispersion, effective mass, etc.\cite{Mona}      

\begin{figure}[t]
\includegraphics[width=0.9\columnwidth]{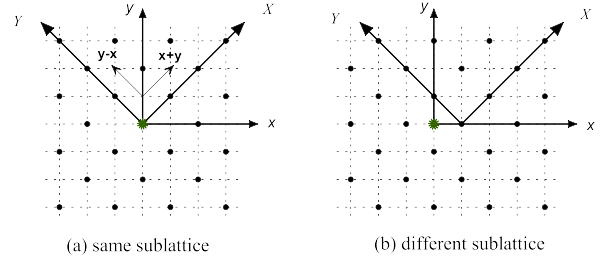}
\caption{(Color online) 
The choice of coordinate systems for the lattice with impurity. The
impurity, shown in green, is at the origin of the $xy$ axes that span the
original lattice with unit vectors ${\bf x, y}$. The $XY$ axes are rotated
by $45\,^{\circ}$ and span the sublattice (black dots) on which the
quasiparticle propagates via the elementary vectors ${\bf y\pm x}$.}  
\label{lattice}
\end{figure}

Equation (\ref{GRG}) can be solved directly in real space by the method of continued
fractions detailed in
Ref. \onlinecite{Mirko}. For completeness, we briefly outline it
here. Let $n$ and $m$ be the $x$ and $y$ components of ${\bf R}\neq
{\bf 0}$ on the coordinates axes $XY$ which is rotated by
$45\,^{\circ}$ with respect to the lattice. It spans the sublattice on
which the quasiparticle moves, shown by the black dots in
Fig. \ref{lattice}(a); its elementary vectors are ${\bf y\pm x}$. In
this coordinate system, Eq. (\ref{GRG}) can be written as 
\begin{multline}
\label{GRG3}
G_{n,m}=\bar{g}_0[-t_1(G_{n+1,m}+G_{n-1,m}+G_{n,m+1}+G_{n,m-1})\\
-t_2(G_{n+1,m+1}+G_{n+1,m-1}+G_{n-1,m+1}+G_{n-1,m-1})],
\end{multline}
where $G_{n,m}\equiv G_{\bf 0,R}(\omega)$ for ${\bf R}=n({\bf y+
  x})+m({\bf y- x})$ is a shorthand notation.  
Eq. (\ref{GRG3}) can be expressed as a single-index recursive relation
by grouping distinct GF's with $n\geq m\geq 0$ into column vectors
$V_M$ according to their Manhattan distance $M=n+m$: 
\begin{equation}
V_{M=2r}=\begin{pmatrix} G_{2r,0} \\ G_{2r-1,1} \\ \vdots\\G_{r,r} \end{pmatrix},  
V_{M=2r-1}=\begin{pmatrix} G_{2r-1,0} \\ G_{2r-2,1} \\ \vdots\\G_{r,r-1} \end{pmatrix}. \nonumber 
\end{equation}
These are the only distinct GFs since all others can be related to
these using symmetries: $G_{n,m}=G_{m,n}=G_{n,-m}=G_{-n,m}$, etc. In
terms of these vectors, Eqs. (\ref{GRG3}) 
be grouped  into the following matrix form 
\begin{equation}
\label{rec1}
\lambda_rV_r=\tilde{\alpha}_rV_{r-2}+\alpha_rV_{r-1}+\beta_rV_{r+1}+\tilde{\beta}_rV_{r+2}
\end{equation}
for $r\geq 2$ and 
\begin{equation}
\label{01}
\begin{array}{rcl}
V_0&=&\bar{g}_0(\omega)+\beta_0V_1+\beta_0V_2 \\
 V_1&=&\alpha_1V_0+\beta_1V_2+\tilde{\beta}_1V_3
 \end{array}
\end{equation} 
for the GFs with $M=0,1$. Here, $\lambda$, $\tilde{\alpha}$, $\alpha$,
$\beta$ and $\tilde{\beta}$ are extremely sparse matrices whose
elements can be read from Eq. (\ref{GRG3}). Combining two copies of
Eq. (\ref{rec1}) corresponding to $r=2s-1$ and $r=2s$ leads to:
\begin{equation}
\label{rec2}
\Gamma_sW_s=A_sW_{s-1}+B_sW_{s+1},
\end{equation} 
where $W_s=\begin{pmatrix} V_{2s-1}\\ V_{2s}\end{pmatrix}$,
$\Gamma_s=\begin{pmatrix} \lambda_{2s-1} & -\beta_{2s-1}
\\ -\alpha_{2s} & \lambda_{2s} \end{pmatrix}$, etc. Because
Eq. (\ref{rec2}) links three consecutive terms, it can be solved in
terms of continued fractions of 
matrices. Specifically, assuming a solution as $W_s=\Omega_sW_{s-1}$ and using it in
Eq. (\ref{rec2}) gives  
\begin{equation}
\Omega_s=(\Gamma_s-B_s\Omega_{s+1})^{-1}A_s,
\end{equation}     
which can be calculated starting from a cutoff $c$ such that
$\Omega_{c+1}=0$. This results in a continued-fraction solution for
$\Omega_s$. In particular, this gives $\Omega_2$ which relates $W_2$
(set of $V_3$ and $V_4$) to $W_1$ (set of $V_1$ and $V_2$). Finally,
the diagonal element of Green's function is found by using these in Eqs. (\ref{01}) and (\ref{rec1}) with $r=2$ to solve for
$V_0=G_{0,0}(\omega)$ from which we find the hole's local density of states
(LDOS):
\begin{align}
\label{figclean}
  \rho({\bf r};\omega)&=-\frac{1}{\pi}\mathrm{Im}\langle 0|h_{\bf
    r}\hat{G}(\omega)h^{\dagger}_{\bf r}|0\rangle\\ 
 &= -\frac{1}{\pi}\mathrm{Im}G_{0,0}(\omega)\nonumber,
\end{align}
which is same as the total density of states in the clean
system. Other GFs $G_{0,{\bf r\ne 0}}(\omega)$ can be then calculated
from $G_{0,0}(\omega)$ using the continued fraction matrices
$\Omega_s$. In practice, the calculation is done on a finite lattice
which is chosen sufficiently large that the GFs become negligible
beyond its boundaries (the broadening $\eta$ introduces an effective
lifetime $1/\eta$ that prevents the quasiparticle from going
arbitrarily far away from its original location). Note that the
equations are modified for the lattice sites close to the boundary: if
the hole can not hop outside the boundary, some of the generalized GS
$F_1, F_2, F_3$ must be set to zero for sites close to the boundary,
resulting in modified effective hoppings $t_1(\omega), t_2(\omega)$,
and on-site energy $\varepsilon(\omega)$ near the boundary. If the cutoff
is large enough, however, the solution becomes insensitive to these
changes. 

\begin{figure}[t]
\includegraphics[width=0.95\columnwidth]{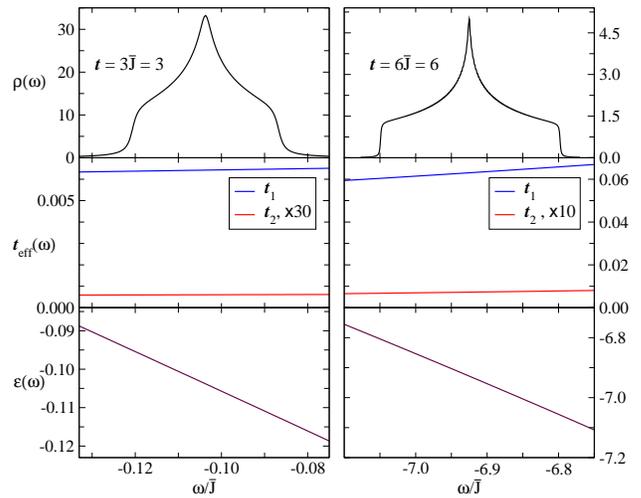}
\caption{(Color online) 
The total density of states (top panels), effective hoppings
$t_1(\omega), t_2(\omega)$ (middle panels) and on-site energy
$\varepsilon(\omega)$ (bottom panels) in the clean system for two
different values of $t/\mathrm{\bar{J}}$. The effective parameters are
relatively constant within the energy band, explaining why the DOS has
the generic form expected for a bare particle with nearest-neighbor
hopping on a square lattice. Here $\mathrm{\bar{J}}=1$, $\eta=10^{-3}$
and $\mathrm{t}=3$ (left panels) and $\mathrm{t}=6$ (right panels),
respectively.}
\label{fig-clean}
\end{figure}

The top panels in Fig. \ref{fig-clean} show the hole's total density
of states (DOS) at two moderate $t/J$ values, for which this
variational approximation was shown to be in good agreement with the
numerical results.\cite{Mona} The quasiparticle bandwidth for $t=6$ is
considerably larger than that for $t=3$, showing the rapid decrease of
the quasiparticle's effective mass with increasing hopping. In the
lower panels we plot the quasiparticle's effective hoppings
$t_1(\omega), t_2(\omega)$ and on-site energy $\varepsilon(\omega)$ over
this energy range. It shows that their energy dependence is relatively
weak in this range and that $t_2(\omega)$, which would make the DOS
asymmetric, is vanishingly small. This explains why the quasiparticle,
in spite of being dressed with magnons, has a DOS similar to that of a
featureless bare particle with only a constant first-nearest-neighbor
hopping.

\section{The effect of the impurity} 
\label{sec:disordered}

In the previous section we confirmed that the hole's motion in the
clean system is described by an effective tight-binding Hamiltonian
with second- and third-nearest-neighbor hoppings which keep the
quasiparticle on the same
sublattice at all times. In this section we investigate the effect
of an attractive impurity on the spectrum of the quasiparticle. The
impurity can be on the  
sublattice in which the quasiparticle moves, or it can be on the other
sublattice. In the former case, one expects the quasiparticle to bind to the
impurity. As mentioned in
the introduction, when they are on different sublattices one might naively
expect the quasiparticle to remain mobile and insensitive to the presence of
impurity. However, we will see that this is not the case.         

\begin{figure}[t]
\includegraphics[width=0.9\columnwidth]{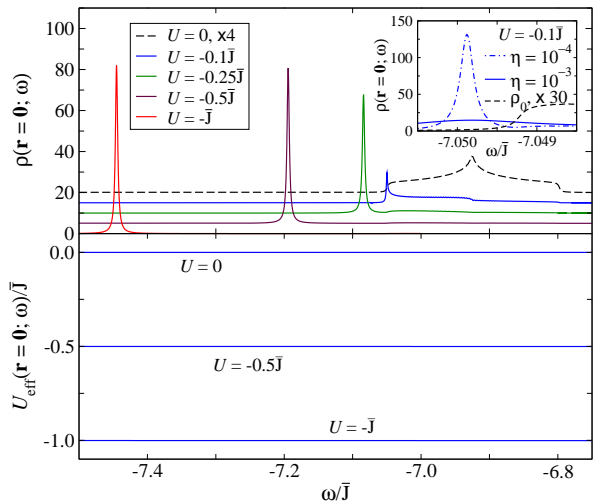}
\caption{(Color online) (Top)
LDOS at the impurity site for various values of $U$. The dashed line is
the DOS in the clean system, times 4. At
finite $U$, a single bound state splits from the continuum and its
binding energy increases with $U$. Curves are shifted vertically to
help visibility. (bottom) The 
effective on-site energy at the impurity site is essentially equal to
$U$. Parameters are $t=6$, $\mathrm{\bar{J}}=1$ and $\eta=10^{-3}$.}  
\label{fig1}
\end{figure}

\subsection{Quasiparticle and impurity on the same sublattice}
The translational invariance of the clean system requires the equal
spreading of the hole's wave function over the entire lattice. This is
expected to change when introducing an attractive impurity and, in
particular, there may exist low-energy bound states where it is
energetically more favourable for the hole to stay close to the
impurity. This tendency can be studied using the hole's Green's
function, $G_{\bf 0, R}(\omega)$, where ${\bf R}$ and the impurity site,
${\bf r=0}$  belong to the same sublattice. This can be calculated
similar to the previous section, while keeping track of the position
of hole with respect to the impurity in order to include the energy
gain $U$ whenever they meet. As a result, some of the
 equations of motion  are changed. For example, Eq. (\ref{GR})
 now reads as
\begin{equation}
\label{GR-U}
G_{\bf 0,R}(\omega)=g_0(\omega; {\bf R})[\delta_{\bf 0,R}-t\sum_{\bf
    u}F_1(\bf R, u;\omega)], 
\end{equation}
where $g_0(\omega; {\bf R})=1/(\omega+i\eta+U\delta_{\bf
  R,0}-4\mathrm{\bar{J}})$. The coefficients
in the equations of motion for $F_2$ also become position-dependent,
reflecting the possibility that the hole is at the impurity
site. The equations for $F_1$ and 
$F_3$,  for which the hole is on the sublattice without the impurity,
remain the same as their counterparts in the clean system.  Tracking
these changed coefficients and their 
effects on the effective hoppings and on-site energies, we now find:
\begin{multline}
\label{GRG2}
G_{\bf 0, R}(\omega)=\tilde{g}_0(\omega;{\bf R})[\delta_{\bf
    R,0}-\sum_{{\bf \boldsymbol\delta}}\tilde{t}_1({\bf R,\boldsymbol
    \delta};\omega)G_{\bf 0, R+{\boldsymbol\delta}}(\omega)\\ 
-\sum_{{\boldsymbol \xi}}\tilde{t}_2({\bf R,\boldsymbol 
  \xi};\omega)G_{\bf 0, R+\boldsymbol\xi}(\omega)], 
\end{multline}
which is similar to Eq. (\ref{GRG}), but now $\tilde{t}_1$ and
$\tilde{t}_2$ depend both on the location and on the direction of
hopping, if {\bf
    R} has the
impurity within the range of its second- or third-nearest-neighbors. If ${\bf R}$ is further away, the effective
parameters take the same values as in the clean system. 
      
Equation (\ref{GRG2}) can be solved similar to Eq. (\ref{GRG}), that is, by
grouping GFs according to their Manhattan distance. Because the
problem has rotational symmetry about the impurity, $G_{\bf
  0,R}(\omega)$ continues to have the same symmetries as in the clean
system, so only the GFs corresponding to $n\geq m\geq 0$ need
to be calculated.

\begin{figure}[t]
\includegraphics[width=0.9\columnwidth]{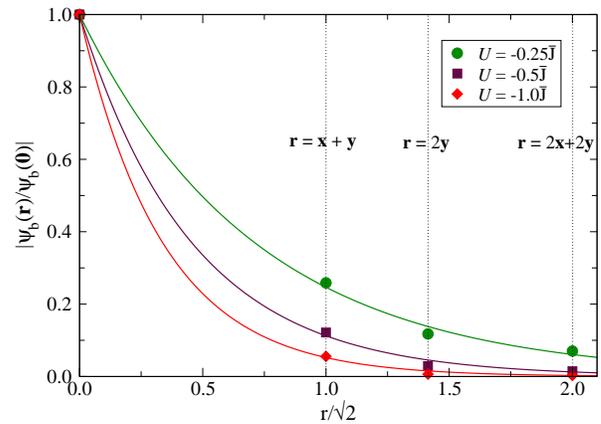}
\caption{(Color online) 
Relative amplitude of the wave functions corresponding to three of the
bound states shown in Fig. \ref{fig1}, at various distances from the
impurity site. Lines are exponential fits. States with bigger
binding energies have shorter decay lengths.}  
\label{decay-sam}
\end{figure}   

Given the almost constant values of $\varepsilon$, $t_1$ , $t_2$ in
this range of energies and
the fact that the problem is two dimensional, bound states are
expected to appear for any finite $U$. The top panel in Fig. \ref{fig1}
shows the LDOS at the impurity site 
${\bf r=0}$ for various values of the on-site attraction $U$. The peaks that
appear below the DOS of the clean system (shown by the dashed line) are 
proportional to Dirac delta functions which are broadened into
Lorentzians by the finite $\eta$. They signal the appearance  of quasiparticle bound
states, characterized by exponential decay of the quasiparticle's wave function
$\psi_b({\bf r})$ away from the impurity. 
The inset 
verifies that this is true even for the smallest $U$: the height of the
"shoulder"-like feature appearing at the bottom of the band in the
main figure scales like $1/\eta$ and evolves into a separate
Lorentzian for small enough $\eta$, showing the presence of a bound
state below the continuum.     

The bottom panel shows the effective attraction at the impurity site
 $U_{\rm eff}({\bf r=0}, \omega) = \mbox{Re}\left[ \varepsilon({\bf
r=0}, \omega )- \varepsilon(\omega)\right]$, i.e. the difference between the
  effective on-site potential at the impurity site and that at sites
  far away from the impurity (or in the clean system). Not surprisingly,
  $U_{\rm eff}({\bf r=0}, \omega) \approx U$, although a small
  dependence of $\omega$ is observed if the scale is significantly expanded.

The exponential decay of the quasiparticle's wave function can be checked explicitly
by calculating the amplitude of these bound states at various
distances ${\bf r}$, which is easily done if we note that at $\omega =
\omega_\mathrm{peak}$ the dominant term in the Lehmann representation is:
\begin{equation}
\label{G-psi}
G_{\bf 0,r}(\omega=\omega_\mathrm{peak})\approx {1\over i
  \eta}\psi_b({\bf 0}) \psi_b({\bf r})^*. 
\end{equation} 

Figure \ref{decay-sam} shows the ratio $|\psi_b({\bf r})/\psi_b({\bf
  0})|= |G_{\bf 0,r}(\omega=\omega_\mathrm{peak})/G_{\bf
  0,0}(\omega=\omega_\mathrm{peak}) |$. The dots are the numerical
values, while the  lines are exponential fits. Those corresponding to
larger binding energies (more negative $\omega_\mathrm{peak}$) are
more tightly bound to the impurity and therefore decay faster, as
expected. This agrees with the larger quasiparticle weight of these states at
${\bf r=0}$, (see Fig. \ref{fig1}).  All these results are quite reasonable.

\begin{figure}[t]
\includegraphics[width=0.9\columnwidth]{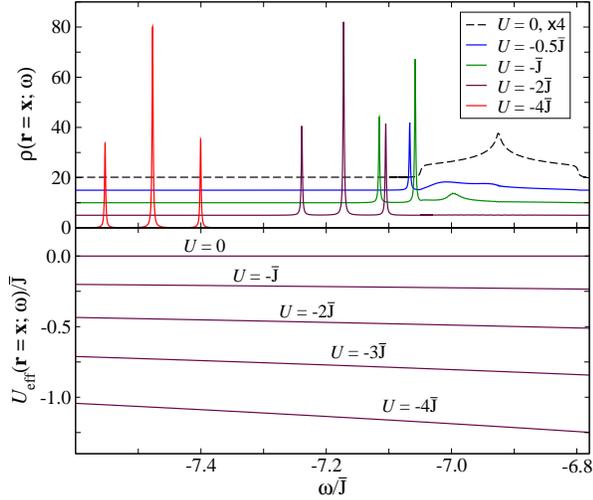}
\caption{(Color online) 
(Top) LDOS $\rho({\bf
  r=x};\omega)$ with curves shifted vertically to help visibility;
  (Bottom) $U_\mathrm{eff}({\bf 
  r=x};\omega)$ at the quasiparticle's sublattice site located nearest
  to the impurity. Up to three bound states 
split from the continuum upon increasing $U$. The presence of the
impurity at ${\bf r=0}$ induces a finite effective on-site attraction at
${\bf r=x}$, whose value is significantly smaller than $U$
(Bottom). 
Parameters are $t=6$, $\mathrm{\bar{J}}=1$, and $\eta=10^{-3}$.}   
\label{fig4}
\end{figure}

\subsection{Quasiparticle and impurity on different sublattices}

We now investigate the more interesting case with the impurity and the quasiparticle
located on different sublattices. To this end, we construct the
Green's function $G_{\bf x,R}(\omega)$ in which ${\bf x}$ and ${\bf R}$
are on the quasiparticle's sublattice [the
rotated frame $XY$ is centered to the right of the
impurity, see Fig. \ref{lattice}(b)]. In particular, we
are interested in the LDOS on this sublattice closest to the impurity $\rho({\bf
  r=x};\omega)=-\mathrm{Im}G_{\bf x,x}(\omega)/\pi$.

The equations of motion for $G_{\bf x,R}(\omega)$  are derived as
before, however now the equations for $F_1$ and $F_3$ are modified by
the presence of the impurity if ${\bf R}$ is close enough to it. This
leads to equations of motion for $G_{\bf x,R}(\omega)$ that are
similar to those in Eq. (\ref{GRG2}), but with different values for
the effective hoppings and on-site energies close to the
impurity. We solve these equations using the same method, but note
that now the number of distinct GFs is higher due to the lower symmetry of
this case.

In Fig. \ref{fig4}, we plot $\rho({\bf r=x};\omega)$ for various values
of $U$. The appearance of Dirac delta peaks shows that bound states
exist in this case as well. A comparison with $\rho({\bf r=0};\omega)$
in Fig. \ref{fig1} for the same value of $U$ shows that these peaks
have different energies, therefore they are distinct states. This is
further confirmed by the fact that up to three bound states appear
here for sufficiently large $U$, as opposed to only one when the
quasiparticle and the impurity were on the same sublattice.

\begin{figure}[t]
\includegraphics[width=0.9\columnwidth]{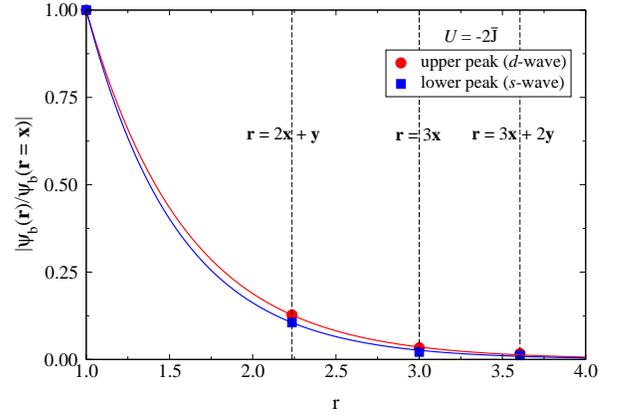}
\caption{(Color online) 
Relative amplitude of the upper and lower 
bound states for $U=-2\mathrm{\bar{J}}$, at various distances from the
impurity site. Lines are exponential fits. }  
\label{decay-dif}
\end{figure} 

These bound states exist in spite of the fact that the impurity is not
located on the sublattice in which the quasiparticle propagates. As
noted above, within a naive picture one does not expect this: the
quasiparticle should not be trapped by an on-site impurity located on
the other sublattice. This shows that the quasiparticle does not
interact with the {\em bare} disorder, but with a {\em renormalized}
one. This comes about because the quasiparticle's effective motion on
one sublattice is made possible via hopping of the hole through the
other sublattice, when the hole and impurity can interact.
Indeed, we define the effective on-site attraction:
$$U_\mathrm{eff}({\bf x};\omega)=\mathrm{Re}[\varepsilon({\bf
    x};\omega)-\varepsilon(\omega)|$$ which again compares the
  effective on-site energy near the impurity to that of sites far away
  from the impurity (or the clean system). This quantity is plotted in
  the lower panel of Fig. \ref{fig4} for various values of $U$. It is
  finite even though the bare disorder at this site is
  zero. $U_\mathrm{eff}({\bf x};\omega)$ is much weaker than
  $U$, as expected since it is an indirect effect; this explains why the
  binding energies for these peaks are much smaller than in the
  previous case. Retardation effects (dependence on $\omega$) are now
  clearly visible, especially for the larger $U$ values. They are due to
  the spin defects accompanying the hole: in order to interact with
  the impurity, the hole must hop onto its sublattice, however its
  ability to do so depends on the structure of the surrounding cloud
  of spin defects. At low energies, the probability for the hole to
  visit the impurity is further suppressed by the energy cost of the
  spin defects generated during hopping, explaining why
  $U_\mathrm{{eff}}$ becomes weaker at these energies. A similar effect 
  has been predicted for  hole-doped CuO ladders with non-magnetic impurities
  that affect the propagating holes even if they do not lie in their path.\cite{Chudzinski}

\begin{figure}[t]
\includegraphics[width=0.9\columnwidth]{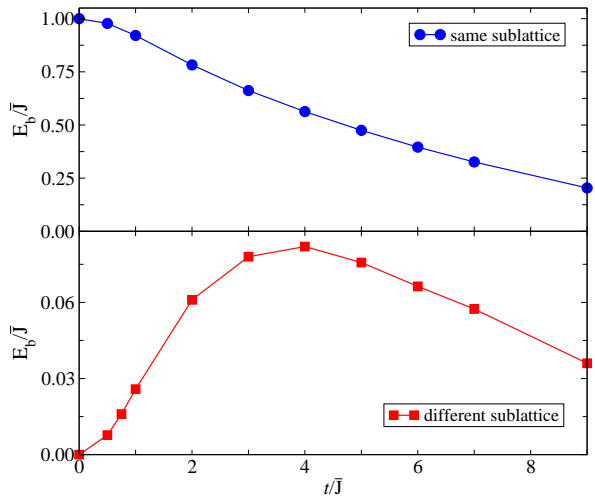}
\caption{Binding energy of the $s$-wave bound state at
  $U=-\mathrm{\bar{J}}$ vs.
  $t/\mathrm{\bar{J}}$, when the quasiparticle and the impurity are on
  the same sublattice (top panel) and different sublattices (bottom panel). 
The smaller binding energy at strong hopping is
  due to the reduction in the quasiparticle's effective mass, which makes it harder
  to trap. When the quasiparticle and the
  impurity are on different sublattices, the enhancement of
  $U_\mathrm{{eff}}$ at small $t$ dominates over the effective
  mass decrease, explaining the growth of the binding energy here.}  
\label{E_b}
\end{figure}

Perturbation theory to zero order in $t$ suggests that there should
be a finite threshold for $U$ in order for bound states to appear. It
can be estimated by comparing the hole's energy at any other site in
the lattice, $4\mathrm{\bar{J}}+ {\cal O}(t^2)$,  to its minimum energy at
the impurity site, $\mathrm{10\bar{J}}-U+  {\cal O}(t^2)$ (the increased energy
is due to the presence of at least one spin defect). If
$U<\mathrm{6\bar{J}}$, this implies that it should not be
energetically favorable for the hole to be at the impurity site. Including
$t^2$ corrections does not change this: a finite threshold value is still
predicted. However, we
do not see any such threshold in the full calculation. This emphasizes
again the importance of the (higher-order) loop 
processes in describing the actual behavior.

As noted, a total of three bound states
emerge upon increasing the impurity attraction $U$. Further increase of
$U$ increases their binding energy, but it does not give rise to more
bound states. One can identify the nature of these bound states by
comparing their amplitudes on the four neighboring sites of the
impurity, $\langle {\bf r=u}|\psi_b\rangle$. These are extracted from
$G_{\bf x,u}(\omega=E_b)$, just as we did in Eq. (\ref{G-psi}). For
the lower peak, we find the same value of $\langle \psi^1_b
|{\bf u}\rangle$ for all ${\bf u}$, implying $s$-wave symmetry. A
state with $s$-wave symmetry is expected to have the strongest binding
to the impurity since, to the leading order in hopping, all of its
four segments meet constructively on the impurity. For the upper peak,
$\langle \psi^3_b  \boldsymbol|{\bf x}\rangle=-\langle \psi^3_b
\boldsymbol|{\bf y}\rangle=\langle \psi^3_b  \boldsymbol|{\bf
  -x}\rangle=-\langle \psi^3_b  \boldsymbol|{\bf -y}\rangle$,
i.e., this state has
$d$-wave symmetry. The middle state has $p_x$ symmetry: $\langle
\psi^2_b  \boldsymbol|{\bf x}\rangle=-\langle \psi^2_b
\boldsymbol|{\bf -x}\rangle$ and $\langle \psi^2_b  \boldsymbol|{\bf
  y}\rangle=\langle \psi^2_b  \boldsymbol|{\bf -y}\rangle=0$. It has a
degenerate twin bound state with $p_y$ symmetry, which has zero
amplitude at ${\bf r=x}$ and therefore it does not appear in $G_{\bf
  x,x}(\omega)$.     
Since the full lattice has rotational symmetry about the impurity, the
resulting bound states are 
expected to mirror this symmetry as well. The spatial profile of
$s$- and $d$-wave states is presented in Fig. \ref{decay-dif}. It
shows that they have very similar decay lengths, consistent with their
fairly similar binding energies and with the
fact that their corresponding peaks in Fig. \ref{fig4} have similar quasi-particle
weights. The $p_x$ state, however, is expected to have about twice
larger weight as it is divided between only the ${\bf x}$ and ${\bf -x}$
lobes, whereas the $s$ and $d$ states have weights equally
distributed in all four directions. Again, this is consistent with its
spectral weight shown in Fig. \ref{fig4}.

\begin{figure}[t]
\includegraphics[width=0.9\columnwidth]{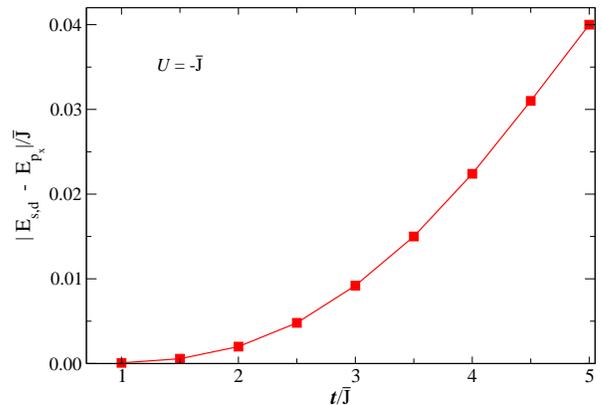}
\caption{The gap between the $p_x$  and either of the $s$ or $d$ states,
  for a fixed $U$. Its enhancement as a function of $t/\mathrm{\bar{J}}$
  reflects the rotational kinetic energy gain of the quasiparticle as
  it becomes lighter with increasing $t$.}  
\label{gap}
\end{figure}                 

Fig. \ref{E_b} shows the hole's binding energy $E_b$ for the
$s$-states as a function of the hopping
$t$, when
$U=-\bar{\mathrm{J}}$. It exhibits quite different trends in
the two cases. As the hopping
becomes stronger, the kinetic energy of the quasiparticle is increased
(its effective mass decreases). A lighter quasiparticle is harder
to trap and this explains why its binding energy at fixed $U$ gets
weaker when it is on the same sublattice with the impurity (top
panel). 

 When the quasiparticle is on the other sublattice (bottom panel) it
 interacts with the impurity by virtue of $U_\mathrm{{eff}}$ which is
 dynamically generated and therefore
 strongly depends on $t$. At $t=0$, the hole is locked at a lattice
 site and is unaware of the presence of impurity, therefore
 $E_b=0$. As $t$ is increased, $U_\mathrm{{eff}}$ is enhanced as the
 hole is able to visit the impurity, whereas the quasiparticle's effective mass
 is reduced as it gains more kinetic energy. The former tends to
 increase the binding energy while the latter reduces it, and
 it is their competition that sets the dependence of binding energy on
 hopping. The initial growth of $E_b$ implies that the enhancement of
 $U_\mathrm{{eff}}$ dominates over the reduction of effective mass at
 small $t$. However, since $U_\mathrm{{eff}}$ is weaker than $U$ for all
 $t$ (Fig. \ref{fig4}), further increase of the hopping makes the hole too
 light to be easily trapped by $U_\mathrm{{eff}}$ and the binding
 energy eventually starts to decrease. While only the
 binding energy of the $s$-wave state is shown in the lower panel
 of Fig. \ref{E_b}, all three peaks exist for small $t$, although they
 are energetically very close to each other. With increasing $t$ they
 move closer to and eventually merge into the continuum such that, at the
 highest $t$ considered in Fig. \ref{E_b}, the $s$-wave state is the only
 existing bound state.  

The energy gaps between the three bound states (when all are present) are
nearly identical. Figure \ref{gap} shows its evolution with $t$ at a
fixed value $U=-\bar{\mathrm{J}}$. Since this must be due to
differences in the rotational kinetic energy, it is expected to
increase with $t$, as the quasiparticle's effective  mass decreases. This
is indeed the observed behavior.

\section{Summary and conclusions}
\label{sec:symmary}

We investigated the effect of a non-magnetic impurity on the motion of
a hole in a 2D square Ising AFM. The
resulting quasiparticle, which  propagates on one
sublattice, is confirmed to form bound states around the
impurity. This is true both when the hole and impurity are on the same
sublattice and when they are on different ones. The latter occurs
because of the renormalization of the effective on-site energy which
results in finite effective attraction at the sites next to the impurity
that can be visited by the quasiparticle. This also explains why a
total of (up to) three bound states with $s$, $p$, and $d$ symmetries
were found in this case, as opposed to only one $s$-wave state in the
case when the quasiparticle is on the same sublattice with the
impurity. In this latter case, the impurity is located in the node of
$p$ and $d$ symmetry states, therefore such states do not see it and
can not be bound to it. (In reality, a non-zero $U_{\rm 
  eff}$ arises at sites different from those occupied by the impurity,
but given the longer distance to the impurity site, this is not large
enough to bind new states). 

Bound states with $s$, $p$, and $d$ symmetries have also been  observed
near an inert vacancy in a Heisenberg AFM. However, in
that case, it is the distortion of the magnetic environment around the
vacancy that binds the hole.\cite{Poilblanc1,Poilblanc2} Such a
distortion is only possible in a Heisenberg model and comes from a
local modification of the spin fluctuations. In an Ising AFM, an
inert site would have no effect on the AFM order of the other
sites. Moreover, if the hole is not allowed to visit this inert
impurity site, there are no Trugman loops including it so the hole
loses kinetic energy when located in that neighborhood. As a result,
we expect that in an Ising AFM, an inert impurity like that of
Ref. \onlinecite{Poilblanc1} would repulse the hole. Bound states
could only appear if a sufficiently strong exchange was turned on
between the hole and the inert spin, so that the exchange energy
gained through it compensated for the loss of kinetic energy. Such a
model was analyzed in 
Ref. \onlinecite{Poilblanc2}, although for the Heisenberg model it was
found that bound states persist only if this exchange with the inert
site is rather weak. All these differences show that the underlying reasons
for the appearance of bound states are very different in these other
models. This is further substantiated by the fact that while a
sublattice dependence is observed in
Refs. \onlinecite{Poilblanc1,Poilblanc2}, it consists of a 
variation of the spectral weight but this is associated with the same
eigenstates. By contrast, in our model, the two sublattices show
different spectra of bound states. 

This result is important because it suggests that two very different
patterns of bound states should be observed with scanning tunneling
microscopy (STM) in such systems, even if only one type of impurity is
present. Note that we assumed that the impurity is located directly at (or
above) a lattice site. If, on the other hand, the impurity was located
either half-way between two sites or in the center of the plaquettes,
then it would not break the symmetry between the two sublattices and
only one pattern of bound states should appear. These 
cases can be studied by
similar means as presented here.

A major simplifying factor of this problem was the assumption of an
Ising AFM. If spin fluctuations are turned on, in a Heisenberg AFM, a
major difference is that the hole no longer needs to go twice around
closed loops in order to become delocalized: spin fluctuations can
remove pairs of neighboring spin defects, thus cutting the string
short and releasing the hole. As a result, one expects a significant
decrease in the effective mass of the quasiparticle, which is indeed
observed.\cite{Trugman} However, it is interesting to note that if
there is true long-range AFM order in the plane (as is the case in
cuprates, due to coupling between planes), the resulting quasiparticle
should continue to primarily reside on one sublattice, because spin
fluctuations can only remove {\em pairs} of spin defects and spin
conservation would continue to make the two sublattices
inequivalent. This suggests that the results we present here, which
are directly traceable to the fact that the quasiparticle lives on one
sublattice, could be relevant for the Heisenberg AFM as well, although
it is impossible to say {\it a priori} if the effective attraction generated
when the quasiparticle and impurity are on different lattices would
suffice to bind states (we would still expect $s$-symmetry bound states to
appear if the quasiparticle and impurity are on the same
sublattice). A follow-up of this issue would be interesting.       

In the broader context, these results confirm the view that coupling
to bosonic degrees of freedom renormalizes not just a
quasiparticle's dispersion, but also the effective disorder it
sees. If the latter 
were not the case, no bound states could arise when the quasiparticle
lives on a difference sublattice than the impurity. Similar large and
non-trivial renormalization of the disorder seen by a dressed
quasiparticle, arising from its coupling to bosons, was also demonstrated
for lattice polarons.\cite{earlier}

\begin{acknowledgments} This work was supported by NSERC,
CIFAR, and QMI. H. E. also acknowledges the UBC Doctoral Four-Year Fellowship award.
\end{acknowledgments}  

\appendix

\section{The equations of motion for $G_{\bf 0,R}(\omega)$}   

Here we present the details of the calculations that lead to
Eq. (\ref{GRG}), which relates the various $G_{\bf 0,R}(\omega)$
GFs. Eq. (\ref{F3}) enables us to eliminate $F_3$ from
Eq. (\ref{F2-3}) to obtain: 
\begin{multline}
\label{F221-1}
F_2({\bf R},{\bf u},{\bf v})-t^2\bar{g}_2g_3F_2({\bf R+u+v, -v, -u})\\
=-t\bar{g}_2F_1({\bf R,u}),
\end{multline}
and
\begin{multline}
\label{F221-2}
F_2({\bf R+u+v},{\bf -v},{\bf -u})-t^2\bar{g}_2g_3F_2({\bf R, u, v})\\
=-t\bar{g}_2F_1({\bf R+u+v,-v}), 
\end{multline}
where $\bar{g}_2=1/(\omega-14\mathrm{\bar{J}}-t^2g_3+i\eta)$ and
Eq. (\ref{F221-2}) results from Eq. (\ref{F221-1}) after changing the
coordinates ${\bf R\rightarrow R+u+v}$, ${\bf u\rightarrow -v}$, ${\bf
  v\rightarrow -u}$. Solving the coupled equations (\ref{F221-1}) and
(\ref{F221-2}), we find 
\begin{multline}
F_2({\bf R},{\bf u},{\bf v})=\gamma_1F_1({\bf R},{\bf u})+\gamma_2F_1({\bf R+u+v},{\bf -v}),
\end{multline} 
in which $\gamma_1=-t\bar{g}_2/[1-(t^2\bar{g}_2g_3)^2]$ and $\gamma_2=t^2\bar{g}_2g_3\gamma_1$. Using this in Eq. (\ref{F1}) gives
\begin{multline}
\label{F1G}
F_1({\bf R},{\bf u})=\\
-t\bar{g}_1[G_{\bf 0, R}+ \gamma_2\sum_{\bf v\perp u} F_1({\bf R+u+v},{\bf -v})],
\end{multline} 
in which $\bar{g}_1=1/(\omega-10\mathrm{\bar{J}}+2t\gamma_1+i\eta)$ and the sum includes the two nearest-neighbor vectors, ${\pm \bf v}$, along the direction perpendicular to ${\bf u}$. With a proper change of coordinates, each $F_1$ on the right-hand side of Eq. (\ref{F1G}) can be expressed in term of a component of $G$ and new $F_1$'s. For example, 
\begin{multline}
\label{F1GF1-1}
F_1({\bf R+u+v},{\bf -v})+t\bar{g}_1G_{\bf0, R+u+v}=\\
-t\bar{g}_1\gamma_2[F_1({\bf R+2u, -u})+F_1({\bf R,u})],
\end{multline}
and
\begin{multline}
\label{F1GF1-2}
F_1({\bf R+u-v},{\bf v})+t\bar{g}_1G_{\bf0, R+u-v}=\\
-t\bar{g}_1\gamma_2[F_1({\bf R+2u, -u})+F_1({\bf R,u})],
\end{multline}
which results after applying either of ${\bf R\rightarrow R+u\pm v}$, ${\bf u\rightarrow \mp v}$, ${\bf v\rightarrow u}$ to Eq. (\ref{F1G}), respectively. 
The additionally introduced $F_1$ can be written in terms of the existing ones by doing ${\bf R\rightarrow R+2u}$, ${\bf u\rightarrow -u}$ on Eq. (\ref{F1G})  
\begin{multline}
\label{F1GF1-3}
F_1({\bf R+2u, -u})+t\bar{g}_1G_{\bf 0, R+2u}=\\
-t\bar{g}_1\gamma_2[F_1({\bf R+u+v, -v})+F_1({\bf R+u-v, v})].
\end{multline}
The four equations (\ref{F1G}) to (\ref{F1GF1-3}) can be
simultaneously solved for the four $F_1$'s in terms of the existing
components of $G$. In particular,  we find:
\begin{multline}
F_1({\bf R,u})=
\zeta_1G_{\bf 0,R}\\
+\zeta_2G_{\bf 0,R+2u}+\zeta_3[G_{\bf R+u+v}+G_{\bf R+u-v}],
\end{multline}
where
$\zeta_1=-t\bar{g}_1[1-2(t\bar{g}_1\gamma_2)^2]/[1-4(t\bar{g}_1\gamma_2)^2]$,
$\zeta_2=-2t\bar{g}_1(t\bar{g}_1\gamma_2)^2/[1-4(t\bar{g}_1\gamma_2)^2]$ 
and $\zeta_3=-t\bar{g}_1\gamma_2(\zeta_1+\zeta_2)$. Finally, using
this in Eq. (\ref{GR}) results in the equation of motion for the GF 
\begin{multline}
\label{GRG-apndx}
G_{\bf 0, R}(\omega)=\bar{g}_0(\omega)[\delta_{\bf R,0}\\
-t_1(\omega)\sum_{{\bf \boldsymbol\delta}}G_{\bf 0,
  R+{\boldsymbol\delta}}(\omega)-t_2(\omega)\sum_{{\boldsymbol
    \xi}}G_{\bf 0, R+\boldsymbol\xi}(\omega)], 
\end{multline}
and its various coefficients are given in the text following
Eq. (\ref{GRG}).  

These effective hoppings and on-site energies are
identical to those  derived for the clean system in
Ref. \onlinecite{Mona}.\cite{typo}  In the presence of disorder, the
solution proceeds similarly but now the various $g$ functions acquire
dependence on the location since their argument is shifted by $U$ if
${\bf R=0}$. This leads to dependence on location (and even
direction of hopping) for the effective hopping and on-site energies,
at sites close enough to the impurity.


\begin{thebibliography}{99}

\bibitem{Kane} C. L. Kane, P. A. Lee, and N. Read, Phys. Rev. B {\bf 39}, 6880 (1989).

\bibitem{Frank} F. Marsiglio, A. E. Ruckenstein, S. Schmitt-Rink, and C. M. Varma, Phys. Rev. B {\bf 43}, 10882 (1991).

\bibitem{Sachdev} S. Sachdev, Phys. Rev. B {\bf 39}, 12232 (1989).

\bibitem{Chubukov} A. V. Chubukov and D. K. Morr, Phys. Rev. B {\bf 57}, 5298 (1998).

\bibitem{Sigia} B. I. Shraiman and E. D. Siggia, Phys. Rev. Lett. {\bf 60}, 740 (1988).

\bibitem{Lee} P. A. Lee, N. Nagaosa, and X. G. Wen, Rev. Mod. Phys. {\bf 78}, 17 (2006).

\bibitem{Phillips} P. Phillips, Rev. Mod. Phys. {\bf 82}, 1719 (2010).

\bibitem{Dagotto} E. Dagotto, R. Joynt, A. Moreo, S. Bacci and E. Gagliano,
Phys. Rev. B {\bf 41}, 9049 (1990); H. Fehske, V. Waas,
H. R{\"o}der and H. B{\"u}ttner, {\it ibid}. {\bf 44}, 8473 (1991); E. Dagotto, Rev. Mod. Phys. {\bf 66}, 763 (1994).

\bibitem{Riera0} J. Riera, S. Koval, D. Poilblanc, and F. Pantigny, Phys. Rev.
B. {\bf 54}, 7441 (1996); G. Xiao, M. Z. Cieplak, J. Q. Xiao, and C. L. Chien, Phys. Rev. B {\bf 42}, 8752 (1990).

\bibitem{basov} D. N. Basov, B. Dabrowski, and T. Timusk, Phys. Rev.
Lett. {\bf 81}, 2132 (1998).  

\bibitem{Alloul} H. Alloul, J. Bobroff, M. Gabay, and P. J. Hirschfeld, Rev. Mod. Phys. {\bf 81}, 45 (2009).

\bibitem{Heidarian} D. Heidarian and N. Trivedi , Phys. Rev.
Lett. {\bf 93}, 126401 (2004); Yun Song, R. Wortis and W. A. Atkinson,
  R. Wortis and W. A. Atkinson, Phys. Rev. B {\bf 77}, 054202 (2008);
Phys. Rev. B {\bf 82}, 073107 (2010).

\bibitem{Suchaneck} A. Suchaneck, V. Hinkov, D. Haug, L. Schulz, C. Bernhard, A. Ivanov, K. Hradil, C. T. Lin, P. Bourges, B. Keimer, and Y. Sidis, Phys. Rev.
Lett. {\bf 105}, 037207 (2010).

\bibitem{wessel} S. Wessel, B. Normand, M. Sigrist, and S. Haas, Phys. Rev.
Lett. {\bf 86}, 1086 (2001).

\bibitem{Poilblanc1} D. Poilblanc, D.J. Scalapino, and W. Hanke, Phys. Rev.
Lett. {\bf 72}, 884 (1994).

\bibitem{Poilblanc2} D. Poilblanc, D.J. Scalapino, and W. Hanke, Phys. Rev.
B. {\bf 50}, 13020 (1994).

\bibitem{Brink} W. F. Brinkman and T. M. Rice, Phys. Rev. B {\bf 2},
  1324 (1970).
  
\bibitem{Poilblanc3} D. Poilblanc, H. J. Schulz, and H. T. Ziman, Phys. Rev. B {\bf 47}, 3268 (1993).

\bibitem{Trugman} S. A. Trugman, Phys. Rev. B {\bf 37}, 1597 (1988).

\bibitem{Mona} M. Berciu and and H. Fehske, Phys. Rev. B {\bf 84},
  165104 (2011). If the constraints are not enforced, this model is
  known as the Edwards model and has been shown to have 
  interesting behavior, for example see: A. Alvermann, D. M. Edwards and
  H. Fehske,  Phys. Rev. Lett. {\bf 98}, 056602 (2007); S. Ejima,
  G. Hager and H. Fehske, Phys. Rev. Lett. {\bf 102}, 106404 (2009).
   
\bibitem{typo} Note that Eq. (18) in Ref. \onlinecite{Mona} has a typo
in the denominator, where the prefactor 4 behind $t_3(\omega)$ has to be replaced with 2. 

\bibitem{Riera} J. Riera and E. Dagotto, Phys. Rev. B {\bf 47}, 15346 (1993); E. Lahoud, O. Nganba Meetei, K. B. Chaska, A. Kanigel, and N. Trivedi,  arXiv:1303.0649v1.

\bibitem{Mirko} M. M\"oller, A. Mukherjee, C. P. J. Adolphs, D. J. J. Marchand, and M. Berciu, J. Phys. A: Math. Theor. {\bf 45}, 115206 (2012).

\bibitem{Chudzinski} P. Chudzinski, M. Gabay, and T. Giamarchi, New J. Phys. {\bf 11}, 055059 (2009).
   
\bibitem{earlier} M. Berciu, A. S. Mishchenko, and N. Nagaosa,
  Europhys. Lett. {\bf 89}, 37007 (2010); H. Ebrahimnejad and
  M. Berciu, Phys. Rev. B {\bf 85}, 165117 (2012); H. Ebrahimnejad and
  M. Berciu, Phys. Rev. B {\bf 86}, 205109 (2012).
 
\end{thebibliography}
\end{document}